\documentclass[conference]{IEEEtran}

\IEEEoverridecommandlockouts
\usepackage{cite}
\usepackage{amsmath,amssymb,amsfonts}
\usepackage{algorithmic}
\usepackage{graphicx}
\usepackage{textcomp}
\usepackage{xcolor}
\def\BibTeX{{\rm B\kern-.05em{\sc i\kern-.025em b}\kern-.08em
    T\kern-.1667em\lower.7ex\hbox{E}\kern-.125emX}}
\begin{document}

\title{A Security-Aware Nonlinearity Study of FPGA-Based Time-to-Digital Converters for Quantum Key Distribution Systems
}




\author{\IEEEauthorblockN{Kun Qin\IEEEauthorrefmark{1}, and Carsten Trinitis\IEEEauthorrefmark{1}}

\IEEEauthorblockA{\IEEEauthorrefmark{1}Chair of Computer Architecture and Operating Systems, Technical University of Munich, Heilbronn, Germany\\
Email: \{kun.qin, carsten.trinitis\}@tum.de}
}

\maketitle


\begin{abstract}
Intrinsic nonlinearity in FPGA-based time-to-digital converters (TDCs) is often treated as a calibration issue and evaluated mainly through post-correction metrics. In quantum key distribution (QKD), however, raw delay-line nonuniformity can affect coincidence timing and thereby influence accidental-coincidence rate and Quantum Bit Error Rate (QBER). This paper analyzes how measured FPGA-TDC nonlinearity propagates to QKD timing metrics using a conservative system-level model that combines random timing uncertainty and deterministic nonlinearity. We also propose fabric-level mitigation strategies based on LUT-assisted delay shaping and placement constraints to reduce severe bin-width irregularities without statistical calibrations. The method is evaluated by reproducing two open-source TDCs implemented on a low-cost Zynq-7000 FPGA. We observe reductions of 14\%-21\% in integral nonlinearity (INL) compared with the non-optimized design, leading to a reduced QBER contribution and an improvement by 3.7\%-14.2\% in the estimated secret fraction. These results suggest that raw FPGA-TDC nonlinearity deserves explicit consideration in timing-sensitive QKD implementations.
\end{abstract}


\section{Introduction}
Quantum key distribution (QKD) is a cryptographic method that distributes secret keys using quantum states (e.g., single photons), where any eavesdropping attempt inevitably disturbs the states and can be detected. In many QKD systems, single-photon detectors (SPDs) are combined with time-taggers. In such systems, the arrival time of photons is required, e.g., to identify coincident photon pairs. The time measurement precision and linearity strongly affect the reliability, including error rate and secure key rate \cite{b1}. QKD evolves from isolated fiber links to multi-node infrastructures and prospective “Quantum Internet of Things” (Q‑IoT) deployments \cite{b2}. End-nodes must perform high-rate, low-jitter time-tagging on compact, power-efficient hardware. Traditionally, this function has been implemented using high-cost ASIC time-taggers \cite{b3}, but there is a strong push towards flexible FPGA solutions that can be embedded at the network edge due to their cost-efficiency, scalability, and flexibility to adapt to various protocols, especially for prototyping small QKD systems \cite{b4}.

Despite substantial progress in FPGA-based TDC design \cite{b5}, imperfect time measurements are often treated primarily as stochastic jitter or as post-calibration issues. In practice, the FPGA delay fabric can introduce both (i) random jitter by manufacturing imperfections and (ii) deterministic nonlinearities from carry-chain delay distortion. These effects can manifest as ultra-wide bins, missing/zero bins, and an imperfect time-to-code transfer function \cite{b6}. However, even when improved linearity is reported, it is typically achieved through statistical averaging and complex calibration techniques, rather than by addressing the physical origins of non-uniformity in the delay-chain structure.  

In contrast, this work investigates how TDC nonlinearity affects QKD timing behavior and derived performance metrics, including the coincidence window and Quantum Bit Error Rate (QBER) \cite{b7}. We also introduce a hardware-level mitigation strategy based on LUT-based delay injection in the FPGA fabric. Our major contributions are:

\begin{itemize}
    \item We derive a conservative analytical model that links measured random timing uncertainty and deterministic TDC nonlinearity to an expansion of the coincidence window and increased QBER contribution.
    \item We propose a fabric-level mitigation method that reduces ultra-wide and near-zero bins by reshaping the delay line in hardware rather than relying only on post-calibration time-to-code correction.
    \item Using raw nonlinearity measurements on a Zynq-7000 platform, we demonstrate reduced TDC nonlinearity after optimization and quantify its implications under representative QKD parameter settings.
\end{itemize}

The remainder of this paper is organized as follows. Section II reviews related calibration strategies and identifies the gap in QKD-oriented nonlinearity analysis. Section III derives the impact of nonlinearity on the coincidence window and QBER. Section IV presents the proposed hardware-level mitigation. Section V reports experimental results and compares them with prior works. Section VI concludes the paper.

\section{Background and Motivation}

\begin{figure}[b]
\centering
\includegraphics[width=\linewidth]{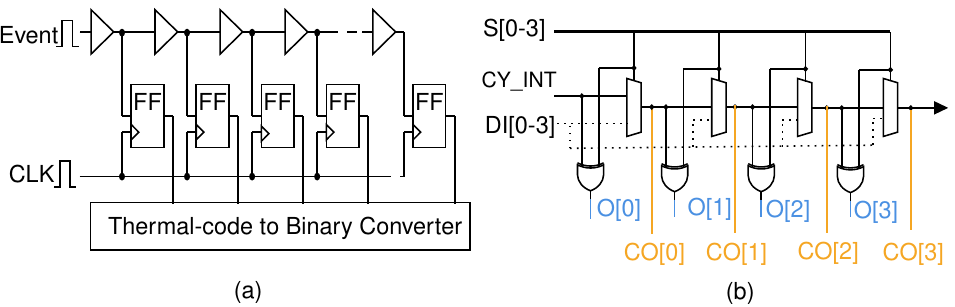}
\caption{Typical structure of a tapped delay line and a CARRY4 primitive. (a) A tapped delay line is formed by a chain of delay elements with Flip-flops for sampling the signal. (b) CARRY4 (Zynq-7000 FPGA) is a dedicated, high-speed carry logic component used to implement fast arithmetic functions, bypassing general-purpose logic fabrics. Each stage, separated by a multiplexer, is used as a delay unit in the tapped delay line.} 
\label{fig:tdl_carry}
\end{figure}
\subsection{Nonlinearity of TDC and Calibration Methods}
FPGA-based TDCs are widely implemented using cascaded CARRY4/8 primitives that form tapped-delay lines (TDLs) \cite{b8, b9, b10}, as illustrated in Fig. \ref{fig:tdl_carry}. They typically require histogram/code-density characterization \cite{b11} to obtain the time-to-code transfer function, from which the single-shot precision (root mean square), bin width (delayed time of one delay cell), the corresponding differential nonlinearity (DNL), and integral nonlinearity (INL) can be extracted. Specifically, DNL is defined by the normalized difference between the estimated bin width and the ideal reference value, and the cumulative deviation of DNL yields the INL, which is the upper-bound deviation of the time measurement. Such histogram-based extraction is effectively unavoidable in practical systems because it provides the empirical mapping from raw codes to time intervals \cite{b12}.


Prior FPGA-based TDC work has explored calibration from several directions. Wu \emph{et al.} combined wave-union sampling with auto-calibration, showing that architectural improvement alone cannot remove bin nonuniformity and drift \cite{b13}. Zhang \emph{et al.} and Parsakordasiabi \emph{et al.} adopted code-density-test-based online calibration with bin-by-bin LUT or RAM correction \cite{b14,b15}. Hua and Chitnis further treated calibration as a state-aware linearization problem by incorporating real propagation states into the encoding and calibration flow \cite{b16}. More recently, Bardpareh \emph{et al.} introduced machine-learning-aided self-calibration as a data-driven alternative to fixed correction tables \cite{b17}. In contrast, Alshahry \emph{et al.} focused mainly on multichannel wave-union design with averaged timestamps  \cite{b18}. These techniques are highly effective when average timing precision after calibration is the main objective. In QKD systems, however, worst-case or strongly nonuniform raw timing distortions can still matter because they may affect the coincidence window required to retain valid events.

\subsection{Security-Aware Coincidence Timing in QKD}
QKD systems rely on accurate identification of correlated detection events to establish a secure key. In this context, the TDC plays a critical role by assigning precise timestamps to detected photons, thereby enabling reliable coincidence matching between the two sides ("Alice" and "Bob"). The coincidence acceptance window then determines the time interval within which two detections are considered a valid coincidence. Enlarging this window can improve the collection of true coincidences, but it also increases accidental coincidences, whose rate is determined by the total singles rate, including both signal-generated singles and dark/invalid counts from single photon detectors. This increases the QBER and, consequently, reduces the security margin available to QKD systems.

\subsection{Gap in QKD-Oriented Analysis of TDC Raw Nonlinearity}

The source-side electronics and the timing backend can be driven from the same reference clock in a QKD system. For example, the laser-driving FPGA and the TDC can share a common external reference \cite{b181}, or the TDC and photon detectors can be locked to the synchronization clock generated by the same source \cite{b182}. In such cases, the arrival phase of detected events may no longer be uniformly distributed with respect to the TDC's clock. Therefore, the measurement can probe only a limited subset of TDC bins repeatedly, so the single-shot precision obtained under random-input assumptions may underestimate the worst-case timing error.

Considering the aforementioned condition, the nonlinearity of TDC is not only a performance factor but also a security-relevant parameter, as it directly influences the selection of the coincidence window and the achievable QBER floor \cite{b21, b22}. For example, although ultra-wide bins can be numerically compensated in the calibrated time-to-code transfer function, they originate from physical nonuniformities in the FPGA carry-chain fabric and are therefore not removed at the hardware level. However, in timing-sensitive QKD systems, such residual timing nonuniformity contributes to the effective timing uncertainty \cite{b6}. It forces a wider coincidence window to avoid rejecting valid photon events. This directly affects the QKD operating point via the elevated error-rate floor (i.e., QBER).

To the best of our knowledge, previous FPGA-based TDC research has predominantly emphasized calibration and post-processing metrics rather than direct physical reshaping of the delay line. Likewise, in QKD systems, while timing jitter is known to influence QBER and secure key rate, the impact of raw TDC nonlinearity prior to calibration on security-relevant behavior has received comparatively little direct attention \cite{b231,b232}.

This work investigates application-specific optimizations of FPGA-based TDCs for QKD systems to mitigate security-relevant timing effects. We further quantify the performance and security benefit of the proposed modifications using the analytical model developed in the next section.

\section{Impact of TDC's Nonlinearity on QKD Metrics}

This section models how TDC's nonlinearity can propagate to QKD-related timing and error metrics: (i) The random single-shot timing uncertainty of the TDC broadens the coincidence-time distribution and thus contributes to the overall system timing jitter. (ii) With our assumption, the deterministic peak-to-peak INL of the TDC is treated as a worst-case timing distortion budget that enlarges the coincidence acceptance window. This separation is motivated by the different physical meanings of stochastic timing uncertainty and deterministic fabric-dependent nonlinearity.

\subsection{TDC's Impact of Random and Deterministic Nonlinearity}

Let $\Delta t_0$ denote the nominal coincidence window in the absence of TDC, and let $W_{\mathrm{INL,pp}}$ denote the peak-to-peak INL interval of the TDC. Under the proposed worst-case assumption, the effective coincidence window is defined as
\begin{equation}
\Delta t_{\mathrm{eff}} = \Delta t_0 + W_{\mathrm{INL,pp}}.
\label{eq:delta_teff_hybrid}
\end{equation}
Note that the expression should be interpreted as an upper-bound approximation: it assumes that the full raw peak-to-peak INL must be tolerated by coincidence logic in the worst case. Then the overall timing jitter can be modeled as
\begin{equation}
\sigma_{\mathrm{sys}}^2 = \sigma_{\mathrm{SPD}}^2 + \sigma_{\mathrm{other}}^2 + \sigma_{\mathrm{TDC}}^2.
\label{eq:sigma_sys_hybrid}
\end{equation}
where $\sigma_{\mathrm{SPD}}$ is the detector timing jitter, $\sigma_{\mathrm{other}}$ collects the remaining non-TDC timing broadening terms, and $\sigma_{\mathrm{TDC}}$ is the single-shot timing uncertainty of the TDC.

Assuming that the coincidence histogram follows a Gaussian distribution with standard deviation $\sigma_{\mathrm{sys}}$, the fraction of true coincidences captured by the effective coincidence window \cite{b24} is
\begin{equation}
\eta_{\mathrm{coin}} = \operatorname{erf} \left( \frac{\Delta t_{\mathrm{eff}}}{2\sqrt{2}\sigma_{\mathrm{sys}}} \right).
\label{eq:eta_coin_hybrid}
\end{equation}

In this formulation, the random TDC uncertainty broadens the coincidence-time distribution through $\sigma_{\mathrm{TDC}}$, thereby reducing the fraction of true coincidences captured within a fixed coincidence window. Meanwhile, the deterministic INL enlarges the required coincidence window through $W_{\mathrm{INL,pp}}$. Although this effect may also increase the captured fraction of true coincidences, its dominant consequence is typically the increase in the accidental coincidence rate \cite{b24}. Therefore, these hypotheses provide a conservative approximation for evaluating how stochastic and deterministic TDC errors may contribute to QBER.

\subsection{TDC-Induced QBER Formulation}

Based on the hybrid timing model above, the QBER contribution associated with the TDC can be quantified through the coincidence detection and accidental coincidence processes.

Let $S_{A,\mathrm{sig}}$ and $S_{B,\mathrm{sig}}$ denote the signal-generated singles rates at Alice (TX) and Bob (RX), respectively, and let $D_A$ and $D_B$ denote the corresponding SPD dark count rates. The total singles rates are therefore
\begin{equation}
S_A = S_{A,\mathrm{sig}} + D_A, \qquad
S_B = S_{B,\mathrm{sig}} + D_B .
\label{eq:singles_total_hybrid}
\end{equation}

Using the standard accidental-coincidence approximation, the accidental coincidence rate \cite{b24} becomes
\begin{equation}
C_{\mathrm{acc}}
\approx
S_A S_B \Delta t_{\mathrm{eff}}
=
\left(S_{A,\mathrm{sig}} + D_A\right)
\left(S_{B,\mathrm{sig}} + D_B\right)
\Delta t_{\mathrm{eff}} .
\label{eq:cacc_hybrid}
\end{equation}

Let $C_{\mathrm{true}}$ denote the true coincidence rate before coincidence-window truncation. The detected coincidence rate is then written in \cite{b24} as
\begin{equation}
C_{\mathrm{det}}
=
\eta_{\mathrm{coin}} C_{\mathrm{true}} + C_{\mathrm{acc}} .
\label{eq:cdet_hybrid}
\end{equation}

Accordingly, the QBER under the proposed hybrid TDC model is derived in \cite{b24} as
\begin{equation}
QBER
=
\frac{
\eta_{\mathrm{coin}} C_{\mathrm{true}} e_{\mathrm{base}}
+
\frac{1}{2} C_{\mathrm{acc}}
}{
\eta_{\mathrm{coin}} C_{\mathrm{true}}
+
C_{\mathrm{acc}}
} .
\label{eq:qber_hybrid}
\end{equation}
where $e_{\mathrm{base}}$ denotes the baseline error contribution from non-TDC impairments, and the factor $1/2$ reflects the fact that accidental coincidences contribute random bits and therefore generate erroneous bits with a probability of 50\% \cite{b24}.

To isolate the TDC-induced QBER penalty, we compare the QBER with and without the TDC contribution. The total incremental TDC-induced QBER is defined as
\begin{equation}
\Delta QBER_{\mathrm{TDC}} = QBER(\sigma_{\mathrm{TDC}}, W_{\mathrm{INL,pp}}) - QBER(0,0) .
\label{eq:delta_qber_tdc_total}
\end{equation}


The proposed model is intended as a conservative system-level approximation. In particular, using $W_{\mathrm{INL,pp}}$ in (1) assumes that the full peak-to-peak raw INL must be tolerated in the worst case by the coincidence-selection logic. This choice may overestimate the required window when only a subset of the raw codes strongly contributes to the distortion. We adopt it here as an upper-bound formulation to make the QKD implication of raw nonlinearity explicit. 

\section{Methodology and Proposed Strategies}

\subsection{Two TDC Designs Under Test}
The proposed mitigation methods are developed by first analyzing the behavior of non-optimized open-source TDC implementations. The extracted raw data shows severe bin nonuniformity and transfer-function discontinuities. It indicates that the dominant impairments originate from fabric-level effects rather than purely random jitter, motivating our mitigation method at the fabric-level of the FPGA delay structure.

\begin{figure}[b]
\centering
\includegraphics[width=\linewidth]{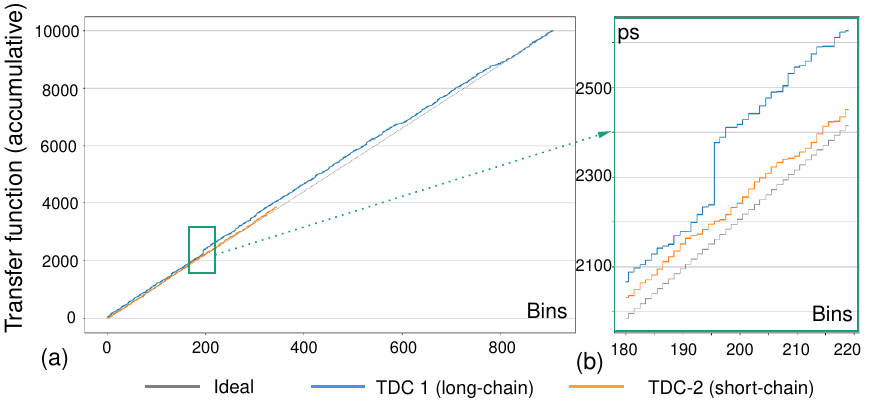}
\caption{Raw bin-to-time transfer functions of TDC-1 and TDC-2. (a) Full delay-chain transfer functions. (b) Zoom-in: a severe local distortion in TDC-1, where an ultra-wide bin produces a step-like offset that propagates to subsequent codes.} 
\label{fig:cdf_tdc1and2}
\end{figure}

To verify that the mitigation is not an artifact of a single delay-line instance or a particular outcome, we reproduce two open-source TDC designs and apply the same optimization flow to both. \textbf{TDC-1} \cite{b25} is a long-chain TDL design (996 bins, 100 MHz) that spans a larger portion of the FPGA fabric and therefore makes clock-region-crossing (CRC) caused clock skewing and routing effects more visible, as shown in Fig. \ref{fig:cdf_tdc1and2}. It is useful for diagnosing the physical origin of deterministic nonlinearity. \textbf{TDC-2} \cite{b26} is a short-chain TDL design (4 $\times$ 96 bins, phase-shifted 260 MHz) that provides a complementary shorter implementation. They are both implemented on Zynq-7000 FPGAs, meeting our low-cost requirement. With them, we demonstrate that our methods remain effective across different delay-line spans and architectures.

\subsection{LUT-Based Delay Injection}
\begin{figure}[b]
\centering
\includegraphics[width=0.8\linewidth, trim= 0 0 130 0, clip]{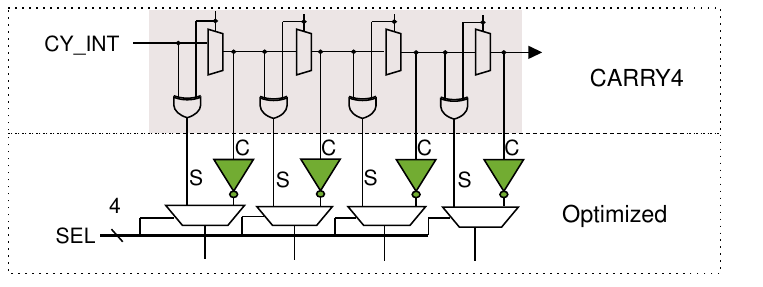}
\caption{Additional LUT-based inverters are added to the delay chain.} 
\label{fig:tdl_tuning}
\end{figure}

The primary mitigation mechanism is a \textbf{LUT-based delay injection} scheme that targets deterministic nonlinearity at its physical origin in the FPGA fabric, as depicted in Fig. \ref{fig:tdl_tuning}. The proposed delay-injection is implemented by introducing inverters and manually selecting the outputs of both "CO" and "C" to achieve the smallest INL, based on observations of the raw output. For example, when a near-zero bin is observed, the sample output can be routed through an LUT-configured inverter before the sampling DFF to introduce additional delay and widen the affected bin. It is employed as a local delay-shaping technique to redistribute the fine-bin widths around problematic regions of the TDL. By decorrelating adjacent tap delays, this method can turn previously unobservable (zero-width) bins into measurable bins in code-density characterization, thereby improving local DNL and overall bin uniformity. Specifically, we insert controlled LUT-induced delays along selected routing paths to (i) reduce extreme delay bins and (ii) smooth abrupt delay variation. Unlike purely statistical post-processing, this approach alters the raw propagation-delay distribution before transfer-function extraction, aiming at reducing severe local bin-width irregularities at the hardware level.

Inserting a LUT route-through between a carry tap and its sampling flip-flop adds delay primarily through the LUT and short local interconnect. By constraining the inverter LUT and the sampling DFF to the same slice (e.g., A6LUT-DFF), the LUT-DFF segment uses the shortest slice-local connection, while the carry chain itself remains on dedicated carry routing along the slice column.

\subsection{Placement-Constrained Delay-Line optimisation}
\begin{figure}[t]
\centering
\includegraphics[width=\linewidth]{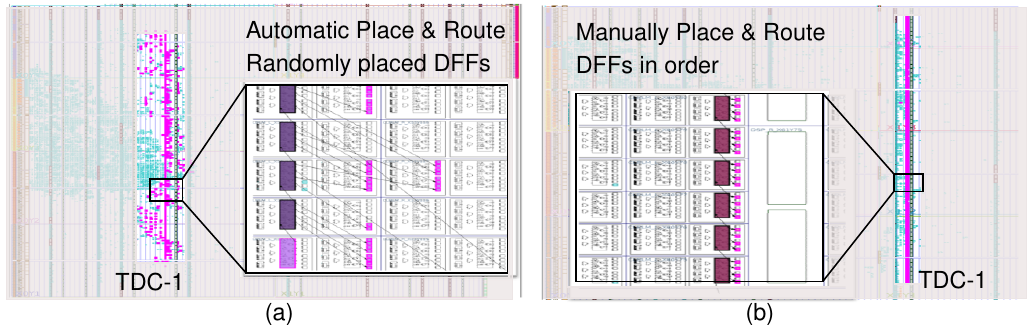}
\caption{Example: Applying LOC constraints to DFFs. } 
\label{fig:placement}
\end{figure}

In this work, manual placement (location constraints for AMD/Xilinx FPGA \cite{b27}) is applied as an \textbf{enhancement step} on top of LUT-based delay injection to further improve delay-line uniformity by suppressing placement-dependent irregularities. Manual placement is well known in FPGA TDC design \cite{b8, b28}, but it is especially important here because the targeted QKD use case is sensitive to large raw bin-width irregularities and deterministic timestamp excursions. This optimisation comprises: (i) keeping the entire delay-chain in the same clock region as much as possible to avoid CRC, (ii) LUT-based inverters are placed in the nearest slices next to the sampling ports, and (iii) DFFs are strictly confined alongside the corresponding delay chain, as shown in Fig. \ref{fig:placement}. By constraining them within a controlled fabric scope, this method improves structural regularity, mitigates CRC-induced artifacts, and enhances the uniformity required for security-sensitive time tagging.

In the present work, optimizations are performed by iteratively inspecting raw transfer-function defects, selecting suspected taps near problematic regions where LUT-based inverters are needed. Placement constraints are applied to confine the delay chain within the same clock region and to spatially align the associated DFFs as a countermeasure for ultra-wide bins.

\section{Results and Discussion}

\subsection{Evaluation Setup}

In our experiment, the reported TDC nonlinearities are extracted from on-board measurements of the Zynq-7020 FPGA, whereas the QBER analysis quantifies the incremental impact of TDC without conducting additional optical experiments. We perform a controlled numerical study based on the timing model in \eqref{eq:delta_teff_hybrid}--\eqref{eq:qber_hybrid}, Section III-B. The TDC-dependent parameters $(W_{\mathrm{INL,pp}}, \sigma_{\mathrm{TDC}})$ are taken from the measured characteristics of the non-optimized and optimized implementations, while detector parameters are set to representative single photon detectors. All non-TDC terms are held constant as in the entanglement-based daylight QKD study \cite{b24} to isolate the incremental contribution of TDC nonlinearity. We sweep the generated singles rates over 0.1 - 2 M counts/s and compute $\Delta QBER_{\mathrm{TDC}}$ by comparing the optimized and non-optimized TDC cases. 

\subsection{Improved Nonlinearity}
We first evaluate the measured raw DNL, INL, and single-shot precision before (Fig. \ref{fig:cdf_tdc1and2}) and after optimization. The baseline design exhibits pronounced nonuniformity, including large peak-to-peak DNL and INL. After applying the proposed mitigation, the raw nonlinearity is visibly reduced, as listed in Tab. \ref{tab:measure_tdc1} and Tab. \ref{tab:measure_tdc2}.


\begin{table}[t]
\caption{TDC-1 before/after the optimization.}
\label{tab:measure_tdc1}
\centering
\setlength{\tabcolsep}{4pt}
\renewcommand{\arraystretch}{1.3 }
 \begin{tabular}{c| c| c | c}
\hline
 \textbf{TDC-1} & \textbf{DNL (ps)}& \textbf{INL (ps)}& \textbf{$\sigma_{\mathbf{TDC}}$} \\
\hline
Raw   & [-11.0, 64.3]  & [-20.0, 280.2] & 14.7 \\
\hline
Optimized   & [-9.3, 20.2]   & [-20.1, 215.7] & 10.9\\
\hline
\% & $\downarrow \mathbf{60}\%$ & $\downarrow \mathbf{21}\%$ & $\downarrow \mathbf{25}\%$ \\
\hline
\end{tabular}
\end{table}

\begin{table}[t]
\caption{TDC-2 before/after the optimization.}
\label{tab:measure_tdc2}
\centering
\setlength{\tabcolsep}{4pt}
\renewcommand{\arraystretch}{1.3}
 \begin{tabular}{c| c| c | c}
\hline
 \textbf{TDC-2} & \textbf{DNL (ps)}& \textbf{INL (ps)}& $\sigma_{\mathbf{TDC}}$ \\
\hline
Raw   & [-8.1, 25.3] & [-35.3, 35.5] & 13.2 \\
\hline
Optimized   & [-8.0, 20.1] & [-29.3, 30.3] & 11.1\\
\hline
\% & $\downarrow \mathbf{16}\%$ & $\downarrow \mathbf{14}\%$ & $\downarrow \mathbf{16}\%$  \\
\hline
\end{tabular}
\end{table}

Across both designs, the proposed mitigation reduces raw deterministic distortion and improves measured single-shot timing precision, though the improvement is greater for the long-chain TDC-1 than for the short-chain TDC-2. The residual peak-to-peak INL in TDC-1 remains large, indicating that the mitigation improves but does not fully eliminate severe long-chain routing artifacts. The possible root cause is that the reduced DNL still accumulates much along the longer delay chain. TDC-2 shows fewer improvements due to its 10x shorter delay chain, and the effect of the routing-level optimization is limited but also effective.

The corresponding measured nonlinearities of the two TDCs under investigation condense the most relevant raw quantities for this paper, without additional calibration steps such as bin-code remapping, averaging, or post-processing of the output code.

\subsection{TDC-induced Incremental QBER Derived by Simulation}
\begin{figure}[t]
\centering
\includegraphics[width=\linewidth]{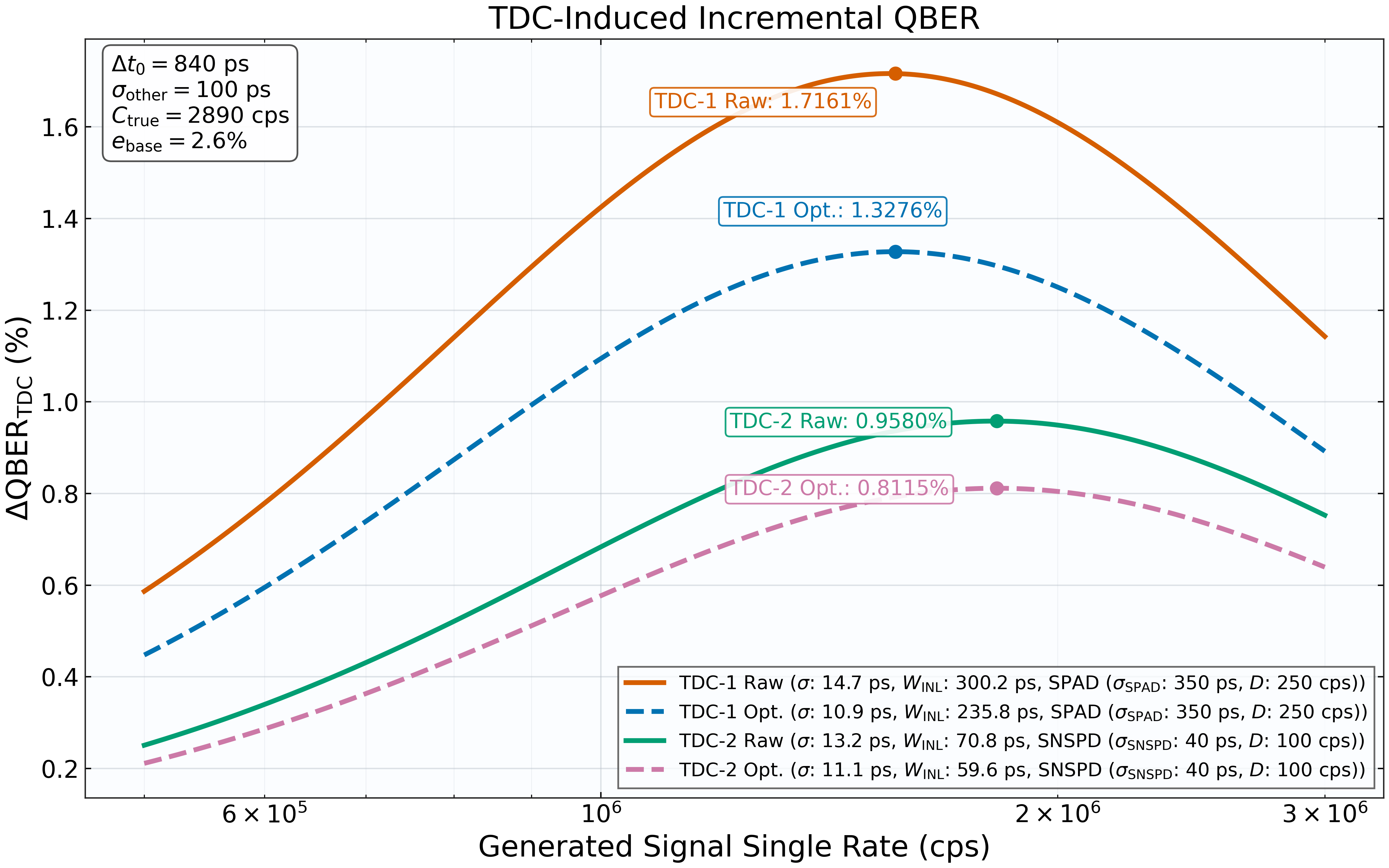}
\caption{Incremental QBER with TDC-1 and SPAD. We sweep the generated signal single rate of around 2 Mcps in our simulation because the observed actual single rate in the experiment was 1.59 Mcps \cite{b22}.}
\label{fig:tdc1_spad}
\end{figure}

Fig. \ref{fig:tdc1_spad} depicts the TDC-induced incremental QBER value for the two TDCs under test. TDC-1 is paired with the single-photon avalanche detector (SPAD), SPCM-800-13 FC from Excelitas\textregistered \cite{b30} with $\sigma=350$ ps and dark count rate $D_{\mathrm{dark}}=250$ counts per second (cps), while keeping the remaining setup parameters fixed according to the daylight QKD study \cite{b24}. Under this configuration, the optimization reduces the peak TDC-induced incremental QBER from 1.71\% to 1.32\% at a signal single rate of 15.5 Mcps, achieving a relative reduction of at most 22.8\%.

For TDC-2, we replace the SPAD with a superconducting nanowire single-photon detector (SNSPD), ID281 from ID Quantique\textregistered \cite{b31}, so that the detector timing uncertainty ($\sigma=40$ ps) is better matched to the INL of TDC-2 ($\approx \pm 30$ ps). This arrangement makes a fair comparison-- if the detector jitter is much larger than the timing uncertainty of TDC under evaluation, the detector rather than the TDC dominates the coincidence-time spread, and the improvement provided by the optimized TDC becomes partially hidden. Besides, this also reflects the practical consideration of properly pairing detectors with TDCs in the system. Under this configuration, the optimization reduces the TDC-induced incremental QBER from 0.95\% to 0.81\% at 18.2 Mcps, corresponding to a 14.7\% relative decrease.



Although the modeled TDC-induced QBER reductions are small in absolute terms, they can still matter when translated through a standard asymptotic BB84 secret-fraction approximation \cite{b29}. Using $r \approx 1 - 2 h_2(QBER)$ only as an illustrative estimate, the observed QBER reductions correspond to non-negligible relative gains in secret fraction. For example, the tiny 0.14\% reduction for TDC-2 lowers the total QBER from 6.77\% to 6.63\%, improving the secret fraction from approximately 0.285 to 0.296, corresponding to a \textbf{3.7\%} relative gain. Optimized TDC-1 achieves \textbf{14.2\%} in the same calculation. These values should be interpreted as indicative to underscore that the raw nonlinearity (without calibration) of TDC has a significant impact on the secret fraction in QKD systems.

\section{Conclusion}

This work studies how raw nonlinearity in FPGA-based TDCs can influence QBER in QKD systems. Using measured data from two TDCs on a Zynq-7020 FPGA, we combine hardware measurements with a conservative analytical timing model to evaluate the impact of random timing uncertainty and deterministic nonlinearity. We further show that LUT-assisted delay shaping can reduce severe raw bin-width irregularities and improve measured timing characteristics. When these measured parameters are propagated through the QKD model, the optimized designs exhibit reduced QBER contribution and improved estimated secret fraction. While the analysis is model-based and does not replace end-to-end optical validation, it suggests that raw FPGA-TDC nonlinearity, without statistical calibration, warrants explicit security consideration in QKD systems. As future work, real free-space QKD experiments can be conducted to investigate the actual impact of TDC nonidealities on QKD performance and security.

\end{document}